\documentclass[
  reprint,
  prd,
  superscriptaddress,
  nofootinbib
]{revtex4-2}

\usepackage{multirow}
\usepackage{slashed}
\usepackage{graphicx}
\usepackage{amsmath}
\usepackage{bm}
\usepackage{yhmath}
\usepackage{hyperref}
\usepackage{subfigure}
\usepackage[T1]{fontenc}
\usepackage{color}
\usepackage{cases}
\usepackage{subfigure}
\usepackage{dcolumn, booktabs, bm}
\usepackage{amsfonts, amssymb, stmaryrd, latexsym, amsmath}
\usepackage{textcomp}
\usepackage{amsthm}
\usepackage{mathrsfs}

\usepackage{array}
\usepackage{caption}
\begin{document}

\title{Chiral mass corrections to decuplet hidden‑charm pentaquark states}

\author{Yu-Quan Guo}
\affiliation{School of Physics, Northwest University, Xi'an 710127, China}
\author{Hao-Song Li}
\email{haosongli@nwu.edu.cn}
\affiliation{School of Physics, Northwest University, Xi'an 710127, China}
\affiliation{Shaanxi Key Laboratory for Theoretical Physics Frontiers, Xi'an 710127, China}
\affiliation{Peng Huanwu Center for Fundamental Theory, Xi'an 710127, China}

\begin{abstract}
	We calculate the masses of the $J^P=\frac{1}{2}^-$ hidden-charm pentaquark states in the $SU(3)$ decuplet within the heavy pentaquark chiral perturbation theory (HPChPT) up to the next-to-leading order (NLO). The one-loop self-energy corrections from the $\pi$, $K$, and $\eta$ mesons are evaluated. Taking $P_\psi^N(4312)$ and $P_{\psi s}^\Lambda(4459)$ as inputs, we predict the masses of the doubly-strange and triply-strange decuplet members $P_{\psi ss}^N$ and $P_{\psi sss}^\Lambda$ to be 4.607 GeV and 4.755 GeV, respectively. The mass splittings between adjacent strangeness layers are nearly constant at approximately 0.148 GeV. This equal-spacing pattern follows directly from the analytical structure of the NLO chiral corrections, while its actual magnitude is set by the low-energy constants (LECs). It indicates that the Gell--Mann--Okubo (GMO) relation, expected at the leading order of SU(3) flavor symmetry breaking, persists to good accuracy at the NLO. 
\end{abstract}
\maketitle

\section{Introduction}

Ever since Gell-Mann and Zweig proposed the quark model, the search for exotic states has been a central pursuit of hadron physics. Conventional hadrons consist of only two types of valence quark configurations: meson $q\bar{q}$ and baryon $qqq$ \cite{GellMann1964,Zweig:1964ruk,Zweig:1964jf}. A hadron is defined as an exotic state when its valence quark composition deviates from the conventional meson $q\bar{q}$ or baryon $qqq$ structures \cite{Wang:2026prg}. Among the many exotic states, pentaquark structures are of significant research value.

In 2015, the LHCb Collaboration reported the first definitive hidden-charm pentaquarks $P^{N}_{\psi}(4380)^{+}$ and $P^{N}_{\psi}(4450)^{+}$, in the decay $\Lambda^{0}_{b}\to J/\psi \, K^{-} p $\cite{LHCb:2015yax}. And in 2019, the LHCb Collaboration not only revealed a new narrow state $P^{N}_{\psi}(4312)^{+}$, but also resolved the $P^{N}_{\psi}(4450)^{+}$ into two distinct peaks, $P^{N}_{\psi}(4440)^{+}$ and $P^{N}_{\psi}(4457)^{+}$. The existence of the $P^{N}_{\psi}(4380)^{+}$ state can neither be confirmed nor excluded \cite{LHCb:2019kea}. A commonly invoked explanation is that they are \(\bar{D}\Sigma_c\) and \(\bar{D}^*\Sigma_c \) bound states \cite{Chen:2019bip,Chen:2019asm,Liu:2018zzu,Liu:2019tjn,Liu:2019tjn,Xiao:2019aya,PavonValderrama:2019nbk,Burns:2019iih,Du:2019pij}, which arises for two reasons. First, the masses of these three pentaquark peaks lie very close to the corresponding baryon-meson two-body thresholds. Second, theoretical predictions of such molecular states had already existed prior to their experimental observation \cite{Wu:2010jy,Wu:2010vk,Wu:2010rv,Yang:2011wz,Xiao:2013yca,Wang:2011rga,Karliner:2015ina}. Although at present the molecular interpretation might be the most favored one \cite{Chen:2019bip,Chen:2019asm,Liu:2018zzu,Liu:2019tjn,Liu:2019tjn,Xiao:2019aya,PavonValderrama:2019nbk,Burns:2019iih,Du:2019pij}, there exist other explanations, e.g., compact pentaquark states \cite{Kuang:2020bnk,Ali:2019npk,Zhu:2019iwm,Wang:2019got,Giron:2019bcs,Cheng:2019obk,Stancu:2019qga}, hadro-charmonia \cite{Eides:2015dtr,Eides:2019tgv,Ferretti:2018ojb}, cusp effects \cite{Kuang:2020bnk}, triangle singularities \cite{Nakamura:2021qvy}, or virtual states \cite{Fernandez-Ramirez:2019koa}. Liu et al. systematically investigated the \(D^{(*)}\Sigma_c^{(*)}\) hadronic molecular system based on LHCb experimental analyses. Besides reproducing the experimentally observed molecular pentaquark states $P^{N}_{\psi}(4312)^{+}$, $P^{N}_{\psi}(4440)^{+}$ and $P^{N}_{\psi}(4457)^{+}$, they additionally predicted four brand-new molecular states, forming a complete heptuplet of molecular pentaquarks satisfying heavy-quark spin symmetry (HQSS) \cite{Liu:2019tjn}. The configuration of this multiplet has also been verified in numerous subsequent independent theoretical studies \cite{Du:2019pij,Xiao:2019aya,Yamaguchi:2019seo,Sakai:2019qph,Xie:2022hhv,PavonValderrama:2019nbk,Meng:2019ilv,Yalikun:2021bfm}. In addition to the \(\bar{D}^{(*)}\Sigma_c\) channels, the \(\bar{D}\Lambda_{c1}(2595)\) channel was claimed to play a nonnegligible role in generating the molecular pentaquark states \cite{Burns:2019iih}, while other studies argue that the contribution of the \(\bar{D}\Lambda_{c1}(2595)\) channel to the formation of molecular states is relatively weak or even negligible \cite{Yalikun:2021bfm}. Various theoretical approaches provide distinct interpretations for the spin-parity quantum numbers of \(P_\psi^N(4440)^+\) and \(P_\psi^N(4457)^+\). Theoretical analyses favor $J^{P}=\frac{1}{2}^{-}$ for $P_\psi^N(4312)^+$ \cite{Du:2019pij}\cite{Lin:2019qiv}. Some studies assign$J^{P}=\frac{1}{2}^{-}$ and $J^{P}=\frac{3}{2}^{-}$ to $P_\psi^N(4440)^+$ and $P_\psi^N(4457)^+$ respectively \cite{Chen:2019asm,Chen:2019bip,Liu:2019tjn,He:2019ify,Xiao:2019mvs,Guo:2019kdc,Xiao:2019aya,Zhang:2019xtu,Wu:2019rog,Wang:2019ato,Xu:2020gjl}, while other quantum number assignment schemes have also been proposed in the literature \cite{Liu:2019tjn,Xu:2020gjl,Zhu:2019iwm,Cheng:2019obk,Yamaguchi:2019seo,Pan:2019skd,Liu:2019zvb}.
 
Subsequently, the LHCb Collaboration reported strange hidden-charm pentaquarks, $P^{\Lambda}_{\psi s}(4338)^{0}$ and $P^{\Lambda}_{\psi s}(4459)^{0}$ \cite{LHCb:2020jpq,LHCb:2022ogu}. Both states lie close to the $\bar{D}\Xi_c$ and $\bar{D}^*\Xi_c$ mass thresholds, respectively, which strongly suggests a hadronic molecular interpretation. Various theoretical approaches, including the one-boson-exchange model \cite{Zhu:2022wpi}, coupled-channel unitary methods \cite{Xiao:2021rgp}, QCD sum rules \cite{Chen:2020uif} and constituent quark models \cite{Ortega:2022uyu}, have studied these two states and confirmed their molecular nature as $\bar{D}^{(*)}\Xi_c$ bound states. The spin-parity quantum number of $P_{\psi s}^{\Lambda}(4338)^0$ has been established as $J^{P}=\frac{1}{2}^{-}$ \cite{LHCb:2022ogu}. Two possible spin-parity assignments, $J^{P}=\frac{1}{2}^{-}$ and $J^{P}=\frac{3}{2}^{-}$, are proposed for $P_{\psi s}^{\Lambda}(4459)^0$ \cite{Zhu:2022wpi}\cite{Xiao:2021rgp}.For convenience, the measured masses and widths of the hidden-charm pentaquark states observed by the LHCb Collaboration, together with their nearest meson-baryon thresholds, are summarized in Table \ref{tab:1}.

\begin{table}[tbp]
	\centering
	\footnotesize
	\caption{Experimental data of the hidden-charm pentaquark states observed by the LHCb Collaboration. The first and second uncertainties are statistical and systematic, respectively.}
	\label{tab:1}
	\begin{tabular}{@{}lccc@{}}
		\toprule
		Pentaquark & Mass (MeV) & Width (MeV) & Threshold \\
		\midrule
		$P_\psi^N(4380)^+$ \cite{LHCb:2015yax}           & $4380\pm8\pm29$                & $205\pm18\pm86$              & $\Sigma_c^*\bar D$ \\
		$P_\psi^N(4312)^+$ \cite{LHCb:2019kea}          & $4311.9\pm0.7_{-0.6}^{+6.8}$   & $9.8\pm2.7_{-4.5}^{+3.7}$    & $\Sigma_c\bar D$ \\
		$P_\psi^N(4440)^+$ \cite{LHCb:2019kea}           & $4440.3\pm1.3_{-4.7}^{+4.1}$   & $20.6\pm4.9_{-10.1}^{+8.7}$  & $\Sigma_c\bar D^*$ \\
		$P_\psi^N(4457)^+$ \cite{LHCb:2019kea}           & $4457.3\pm0.6_{-1.7}^{+4.1}$   & $6.4\pm2.0_{-1.9}^{+5.7}$    & $\Sigma_c\bar D^*$ \\
		$P_{\psi s}^\Lambda(4459)^0$ \cite{LHCb:2020jpq} & $4458.8\pm2.9_{-1.1}^{+4.7}$   & $17.3\pm6.5_{-5.7}^{+8.0}$   & $\Xi_c\bar D^*$ \\
		$P_{\psi s}^\Lambda(4338)^0$ \cite{LHCb:2022ogu} & $4338.2\pm0.7\pm0.4$           & $7.0\pm1.2\pm1.3$            & $\Xi_c\bar D\,(\tfrac12^-)$ \\
		\bottomrule
	\end{tabular}
\end{table}

	Quantum chromodynamics (QCD) is the fundamental gauge theory of strong interactions. Thanks to asymptotic freedom, perturbative QCD works well at high energies, yet it fails in low-energy regions: the coupling constant becomes large, and we cannot analytically compute low-energy hadron properties from first principles. A key feature of QCD is spontaneous chiral symmetry breaking, which produces light pseudoscalar Goldstone bosons (\(\pi, K, \eta\)). Chiral perturbation theory (ChPT) is the effective field theory built on QCD’s chiral symmetry, describing low-energy Goldstone-boson dynamics. Following Weinberg’s power counting, it expands the chiral-invariant effective Lagrangian in powers of momenta and quark masses, absorbing loop divergences via renormalizable LECs. ChPT has two main branches: mesonic ChPT and baryonic ChPT. The mesonic sector includes the leading-order \(\mathcal{L}_2\), the \(\mathcal{O}(p^4)\) Gasser–Leutwyler Lagrangian, and the Wess–Zumino–Witten action for chiral anomalies. For baryons, the heavy-baryon scheme solves power-counting problems in relativistic calculations. As a standard low-energy theoretical tool, ChPT accurately predicts the masses, scattering processes and various electromagnetic observables of light hadrons \cite{Scherer:2002tk}.

Inspired by the heavy-baryon chiral perturbation theory, we adopt a similar heavy-particle formalism and develop the HPChPT, where pentaquark states are treated as heavy static sources \cite{Li:2024jlq}.

Within the molecular picture, hidden-charm molecular pentaquarks consist of a singly charmed baryon and an anti-charmed meson. Under SU(3) flavor symmetry, singly charmed baryons are obtained by substituting one light quark of conventional light baryons with a charm quark. The two light quarks inside the singly charmed baryon can form either symmetric or antisymmetric pairs.

If the light quark pair is symmetric, the singly charmed baryon transforms as the $\boldsymbol{6}_f$ representation. Coupled with the anti-charmed meson belonging to the $\boldsymbol{\bar{3}}_f$ representation, the composite system produces pentaquarks in the $\boldsymbol{10}_f$ decuplet and $\boldsymbol{8}_{1f}$ octet representations. If the light quark pair is antisymmetric, the singly charmed baryon belongs to the $\boldsymbol{\bar{3}}_f$ representation, and the coupling with the anti-charmed meson generates pentaquarks of the $\boldsymbol{8}_{2f}$ octet and $\boldsymbol{1}_f$ singlet. The corresponding group-theoretical decomposition reads
\begin{equation}
	\boldsymbol{3}\otimes\boldsymbol{3}\otimes\boldsymbol{3} = \boldsymbol{1}\oplus\boldsymbol{8}_1\oplus\boldsymbol{8}_2\oplus\boldsymbol{10}.\label{eq:1}
\end{equation}

\begin{figure}[htbp]
	\centering
	\subfigure[Flavor decuplet $\boldsymbol{10}_f$]{
		\includegraphics[width=0.48\textwidth]{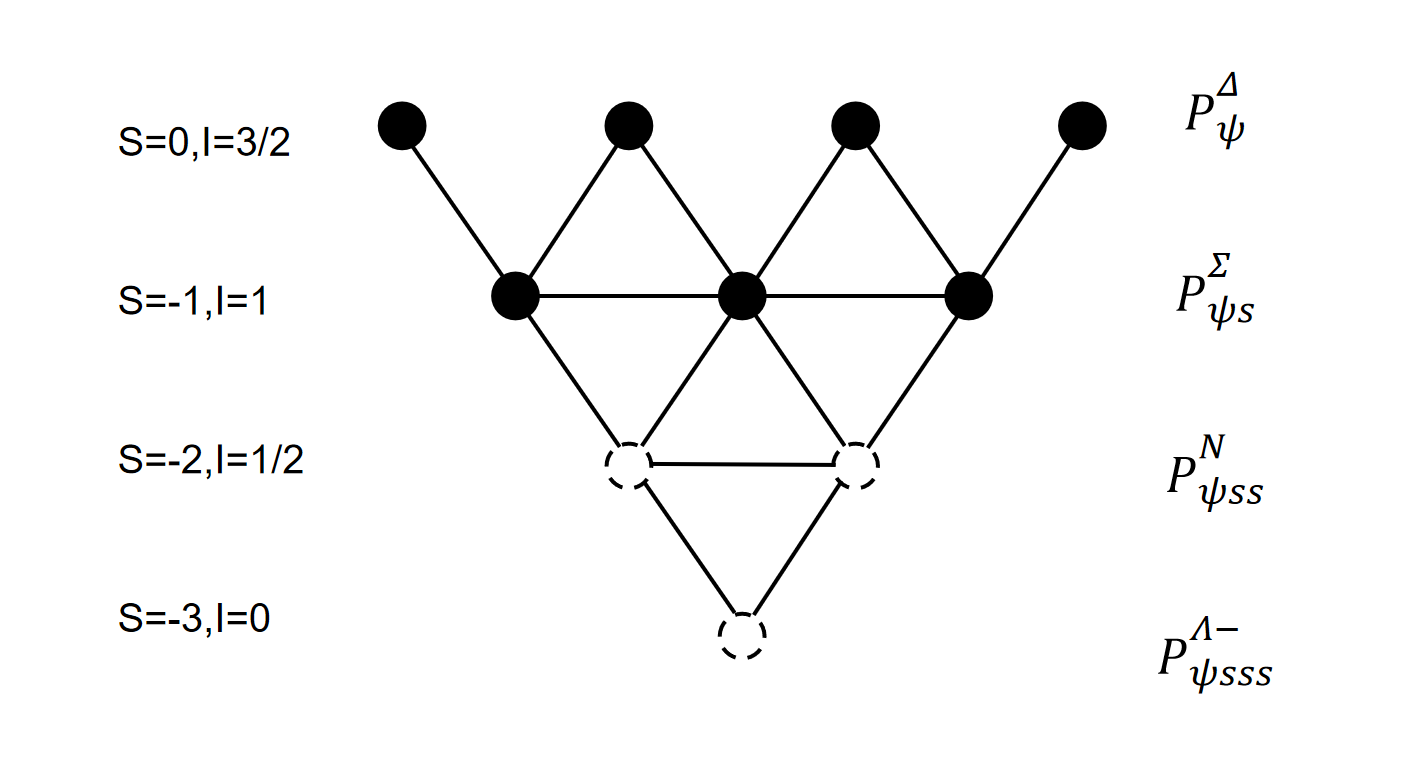}}
	\hfill
	\subfigure[Flavor octet $\boldsymbol{8}_{f}$]{
		\includegraphics[width=0.48\textwidth]{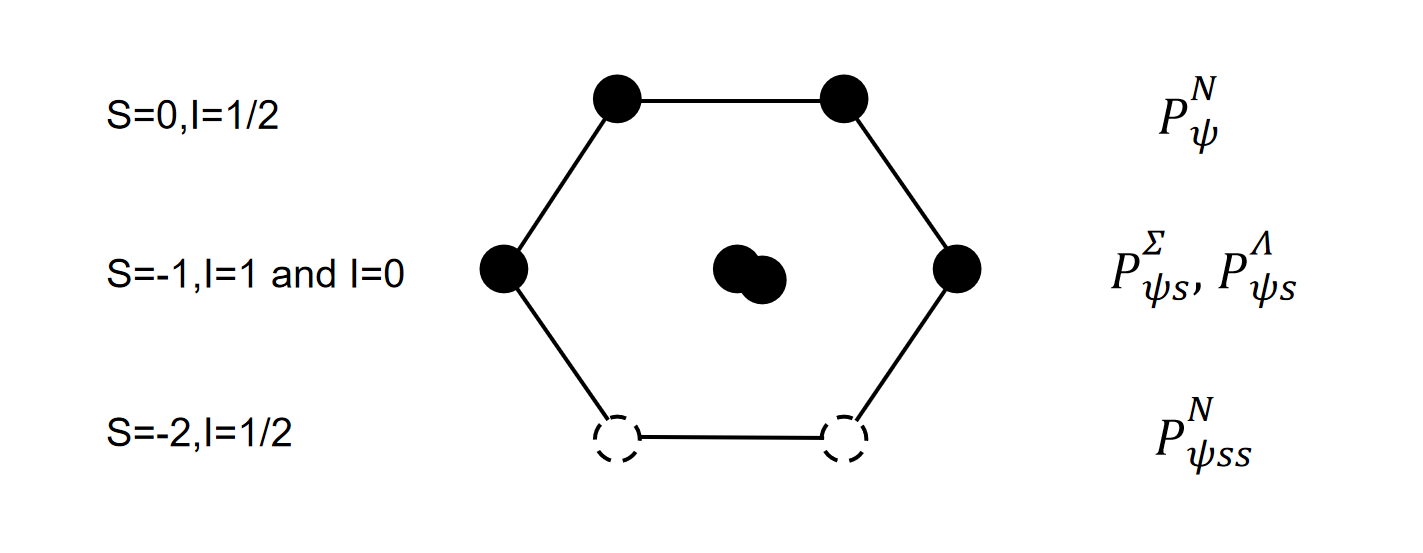}}
	\caption{The flavor contents of the two multiplets produced by the symmetric
		configuration. (a) The
		fully symmetric decuplet $\boldsymbol{10}_f$, with $4+3+2+1$ members
		along the strangeness layers $S=0,-1,-2,-3$ and isospin
		$I=3/2,1,1/2,0$. (b) The companion
		octet $\boldsymbol{8}_{f}$ on a hexagonal lattice; at $S=-1$ the $I=1$
		triplet and the $I=0$ singlet coincide at $I_3=0$ and are shown as
		partially overlapping circles. Filled circles represent states with
		experimental candidates, while dashed open circles denote predicted
		high-strangeness pentaquarks.}
	\label{fig:1}
\end{figure}

The flavor contents of the two multiplets produced by the symmetric
configuration, namely the decuplet $\boldsymbol{10}_f$ and the octet
$\boldsymbol{8}_{f}$, are visualized in the $I_3$--strangeness weight
diagrams shown in Fig. \ref{fig:1}.

As shown in \ref{fig:1}(a), the decuplet is described by a
fully symmetric rank-three flavor tensor and contains ten members arranged
in successive strangeness layers: an $I=3/2$ quartet at $S=0$, an $I=1$
triplet at $S=-1$, an $I=1/2$ doublet at $S=-2$, and an $I=0$ singlet at
$S=-3$.

The companion octet $\boldsymbol{8}_{f}$, illustrated in
 \ref{fig:1}(b), comprises eight states on a hexagonal lattice.
At $S=0$ it contains an $I=1/2$ doublet. At $S=-1$ it accommodates two
distinct isospin multiplets, an $I=1$ triplet $P_{\psi s}^{\Sigma}$ and
an $I=0$ singlet $P_{\psi s}^{\Lambda}$; they share the same $I_3=0$
coordinate and are therefore represented by partially overlapping circles.
At $S=-2$ there appears another $I=1/2$ doublet. 

In both panels, filled circles denote the states already observed or
having experimental candidates, while dashed open circles stand for the
higher-strangeness pentaquarks predicted by SU(3) symmetry but not yet
experimentally established.

In our previous work, we mainly focused on the SU(3) flavor octet multiplets of hidden-charm molecular pentaquarks. Within the HPChPT framework, we have systematically studied the chiral mass corrections for both $J^P = \frac{1}{2}^-$ and $J^P = \frac{3}{2}^-$ octet hidden-charm pentaquarks \cite{Li:2025hpd} \cite{Li:2025vjw}. Having investigated the octet multiplets, it is natural to further consider the decuplet counterparts. Inspired by our previous calculations, we generalize this formalism to the $J^P = \frac{1}{2}^-$ decuplet pentaquark system in the present work. The decuplet pentaquarks feature chiral interaction vertices different from those of the spin-$1/2$ octet case. Following the heavy-particle formalism of heavy-baryon chiral perturbation theory, we treat decuplet pentaquarks as heavy static sources and construct the SU(3) chiral effective Lagrangian. We evaluate both the tree-level mass terms and the next-to-leading order mass shifts originating from pseudoscalar meson one-loop self-energy diagrams, and predict the SU(3) flavor-breaking mass splittings inside the decuplet multiplet. 

Section II is devoted to the analytical derivation of the NLO mass corrections for the decuplet pentaquarks. In Section III, we present the numerical results, examine the equal-spacing pattern of the mass splittings, and discuss the physical implications. A summary and outlook are given in Section IV.

\section{Analytical Derivation of the Decuplet Pentaquark Masses}
To calculate the masses of spin-$1/2$ decuplet pentaquark states, we construct the free Lagrangian for pentaquark states, expressed as:
\begin{equation}
	\mathcal{L}_0^{(1)}=\mathrm{Tr}\big[\bar{P}(iv\cdot D)P\big], \label{eq:2}
\end{equation} 
In Eq (\ref{eq:2}), $P$ represents the decuplet hidden-charm molecular pentaquark field, $v^\mu$ is the constant four-velocity of the heavy pentaquark satisfying $v^2=1$, and $D_\mu$ is the chiral covariant derivative, with $v\cdot D\equiv v^\mu D_\mu$. The related definitions are presented below.

We employ the representation of $P$ given in Ref.~\cite{Li:2025ddx}, and its components are defined as follows:
\begin{align}
	&P^{111}=P_\psi^{\Delta^{++}},\quad P^{112}=\frac{1}{\sqrt{3}}P_\psi^{\Delta^{+}},\quad P^{122}=\frac{1}{\sqrt{3}}P_\psi^{\Delta^{0}},\nonumber\\
	&P^{222}=P_\psi^{\Delta^{-}},\quad P^{113}=\frac{1}{\sqrt{3}}P_{\psi s}^{\Sigma^{+}},\quad P^{123}=\frac{1}{\sqrt{6}}P_{\psi s}^{\Sigma^{0}},\nonumber\\
	&P^{223}=\frac{1}{\sqrt{3}}P_{\psi s}^{\Sigma^{-}},\quad P^{133}=\frac{1}{\sqrt{3}}P_{\psi ss}^{N^{0}},\quad P^{233}=\frac{1}{\sqrt{3}}P_{\psi ss}^{N^{-}},\nonumber\\
	&P^{333}=P_{\psi sss}^{\Lambda^{-}}. \label{eq:3}
\end{align}
The covariant derivative acting on the pentaquark state is defined as
\begin{align}
	D_\mu &= \partial_\mu + \frac{1}{2}\big(u^\dagger\partial_\mu u + u\partial_\mu u^\dagger\big), \label{eq:4}
\end{align}
\begin{align}
	u^2 &= U = \exp(i\phi/F_\phi), \label{eq:5}
\end{align}
$\phi$ denotes the pseudoscalar meson field, whose explicit matrix form reads
\begin{equation}
	\phi=
	\begin{pmatrix}
		\pi^0+\dfrac{1}{\sqrt{3}}\eta & \sqrt{2}\pi^+ & \sqrt{2}K^+\\
		\sqrt{2}\pi^- & -\pi^0+\dfrac{1}{\sqrt{3}}\eta & \sqrt{2}K^0\\
		\sqrt{2}K^- & \sqrt{2}\bar{K}^0 & -\dfrac{2}{\sqrt{3}}\eta
	\end{pmatrix}. 
	\label{eq:6} 
\end{equation}
where $F_\phi$ is the Goldstone boson decay constant, $F_\pi\approx0.924$ GeV, $F_K\approx0.113$ GeV and $F_\eta\approx0.116$ GeV \cite{Scherer:2002tk}.

The Lagrangian contributing to the masses of decuplet pentaquarks at leading order reads \cite{Li:2025ddx}
\begin{equation}
\
\mathcal{L}_{\mathrm{int}}^{\frac{1}{2}}=\mathrm{Tr}\big(-2d_1\bar{P}S_\mu u^\mu P\big),
\label{eq:7}
\end{equation}
where $d_1$ is a coupling constant and $S_\mu$ is a spin operator defined via
\begin{equation}
	S_\mu=\frac{i}{2}\gamma_5\sigma^{\mu\nu}v_\nu
	=-\frac{1}{2}\gamma_5\big(\gamma^\mu\slashed{v}-v^\mu\big),
	\label{eq:8}
\end{equation}
with $\gamma_5$ and $\sigma^{\mu\nu}$ the standard Dirac matrices. The axial‑vector field $u_\mu$ is defined as
\begin{equation}
	u_{\mu}=\frac{i}{2}\big[u^{\dagger}\partial_{\mu}u-u\partial_{\mu}u^{\dagger}\big]. \label{eq:9}
\end{equation}

The leading-order chiral effective Lagrangians responsible for tree-level mass corrections are also needed
\begin{eqnarray}
\mathcal{L}_{\mathrm{tree}}^{(2)}=f_1\bar{P}\chi_{+}{P}, \label{eq:10}
\end{eqnarray}
where $f_1$ is the low-energy constant. The chiral block \(\chi_+\) is defined as
\begin{eqnarray}
	\chi_{+}=u^{\dagger}\chi u^{\dagger}+u\chi^{\dagger}u, \label{eq:11}
\end{eqnarray}
\(\chi\) is defined as
\begin{eqnarray}
\chi=\left(\begin{array}{ccc}
	M^{2}_{\pi} & 0 & 0\\
		0 & M^{2}_{\pi} & 0\\
		0 & 0 & 2M^{2}_{K}-M^{2}_{\pi}
\end{array}\right). \label{eq:12}
\end{eqnarray}
The pseudoscalar meson masses are set to \(M_\pi=0.140\ \mathrm{GeV}\), \(M_K=0.494\ \mathrm{GeV}\), \(M_\eta=0.550\ \mathrm{GeV}\).

The pentaquark states are considered as heavy sources, and equivalently their momentum decomposes as
\begin{eqnarray}
	 p_{\mu} = m_{0}v_{\mu} + k_{\mu}. \label{eq:13}
\end{eqnarray}
where \(m_0\) denotes the pentaquark mass in the chiral limit, and \(k_\mu\) stands for a small off-shell momentum satisfying \(v\cdot k \ll m_0\).

\begin{figure}[htbp]
\centering
\includegraphics[width=1.0\linewidth]{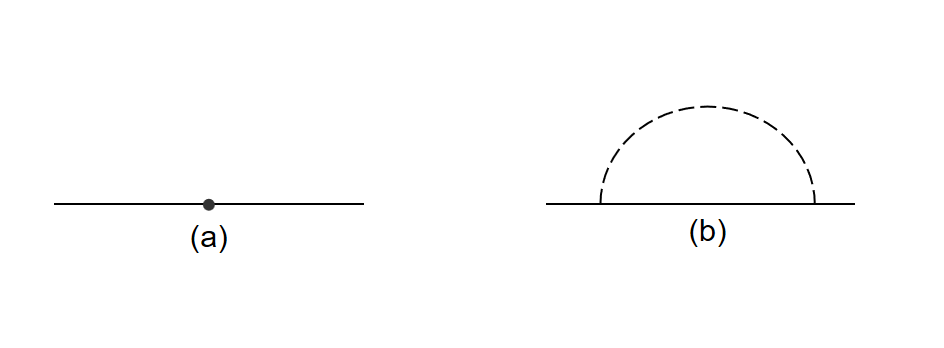}
\caption{Feynman diagrams contributing to the mass corrections of decuplet pentaquark states up to NLO. The solid lines and dashed lines stand for the decuplet pentaquark states and the Goldstone boson, respectively. Solid black dots denote second-order couplings. }\label{fig:2}
\end{figure}

Fig. \ref{fig:2} shows the one-particle irreducible Feynman diagrams that contribute to mass corrections for decuplet pentaquarks with spin-parity \(J^P = \frac{1}{2}^-\).

The full propagator is written as
\begin{eqnarray}
	iG(\omega) = \frac{i}{v\cdot p - m_0 - \Sigma_{P_D}(\omega)},\label{eq:14}
\end{eqnarray}
where \(\omega = v\cdot k\). \(\Sigma_{P_D}(\omega)\)denotes the higher‑order correction to the self‑energy. The physical pentaquark mass \(m_{P_D}\) corresponds to the pole of the propagator with \(p=m_{P_D}v\), 
\begin{eqnarray}
m_{P_D} = m_0 + \Sigma_{P_D}(0)\label{eq:15}.
\end{eqnarray}
At order \(O(p^2)\), the tree-level contribution corresponding to Fig. \ref{fig:2}(a) can be calculated from the chiral symmetry breaking Lagrangians in Eq. (\ref{eq:10}). At order \(O(p^3)\), the loop diagram in Fig. \ref{fig:2}(b) contributes to the self-energy,
\begin{eqnarray}
	\Sigma_{P_D} = -C^{loop}_{\phi}\frac{M^{3}_{\phi}}{2(8F_{\phi})^{2}\pi},\label{eq:16}
\end{eqnarray}
where the coefficients $C^{loop}_{\phi}$ are given in Table \ref{tab:2}.
\begin{table}[htp]
	\centering
	\caption{Coefficients $C^{loop}_{\phi}$}
	\vspace{0.1em}
	\label{tab:2}
	\renewcommand{\arraystretch}{2.5}
	\begin{tabular}{c|c|c|c|c}
		\toprule[1.0pt]
		\toprule[1.0pt]
		States & $P_{\psi}^{\Delta}$ & $P_{\psi s}^{\Sigma}$ & $P_{\psi ss}^{N}$ & $P_{\psi sss}^{\Lambda}$ \\
		\hline
		$C^{loop}_{\pi}$ & $\textstyle\frac{20}{3}d^2_{1}$ & $\textstyle\frac{32}{9}d^2_{1}$ & $\textstyle\frac{4}{3}d^2_{1}$ & $0$ \\
		\hline
		$C^{loop}_{K}$ & $\textstyle\frac{8}{3}d^2_{1}$ & $\textstyle\frac{64}{9}d^2_{1}$ & $8d^2_{1}$ & $\textstyle\frac{16}{3}d^2_{1}$ \\
		\hline
		$C^{loop}_{\eta}$ & $\textstyle\frac{4}{3}d^2_{1}$ & $0$ & $\textstyle\frac{4}{3}d^2_{1}$ & $\textstyle\frac{16}{3}d^2_{1}$ \\
		\toprule[1.0pt]
		\toprule[1.0pt]
	\end{tabular}
\end{table}

The physical masses of decuplet pentaquarks up to NLO are expressed explicitly as,
\begin{eqnarray}
	m_{P_{\psi}^{\Delta}}&=&m_{0}-2f_1M^{2}_{\pi}\nonumber\\
	&&-\frac{20}{3}d^{2}_{1}\frac{M^{3}_{\pi}}{2(8F_{\pi})^{2}\pi}-\frac{8}{3}d^{2}_{1}\frac{M^{3}_{K}}{2(8F_{K})^{2}\pi}\nonumber\\
	&&-\frac{4}{3}d^{2}_{1}\frac{M^{3}_{\eta}}{2(8F_{\eta})^{2}\pi},\label{eq:17}
\end{eqnarray}
\begin{eqnarray}
	m_{P_{\psi s}^{\Sigma}}&=&m_{0}-\frac{2}{3}f_1(2M^{2}_{K}+M^{2}_{\pi})\nonumber\\
	&&-\frac{32}{9}d^{2}_{1}\frac{M^{3}_{\pi}}{2(8F_{\pi})^{2}\pi}-\frac{64}{9}d^{2}_{1}\frac{M^{3}_{K}}{2(8F_{K})^{2}\pi},\label{eq:18}
\end{eqnarray}
\begin{eqnarray}
	m_{P_{\psi ss }^{N}}&=&m_{0}-\frac{2}{3}f_1(4M^{2}_{K}-M^{2}_{\pi})\nonumber\\
	&&-\frac{4}{3}d^{2}_{1}\frac{M^{3}_{\pi}}{2(8F_{\pi})^{2}\pi}-8d^{2}_{1}\frac{M^{3}_{K}}{2(8F_{K})^{2}\pi}\nonumber\\
	&&-\frac{4}{3}d^{2}_{1}\frac{M^{3}_{\eta}}{2(8F_{\eta})^{2}\pi},\label{eq:19}
\end{eqnarray}
\begin{eqnarray}
	m_{P_{\psi sss}^{\Lambda}}&=&m_{0}-\frac{2}{3}f_1(6M^{2}_{K}-3M^{2}_{\pi})\nonumber\\
	&&-\frac{16}{3}d^{2}_{1}\frac{M^{3}_{K}}{2(8F_{K})^{2}\pi}-\frac{16}{3}d^{2}_{1}\frac{M^{3}_{\eta}}{2(8F_{\eta})^{2}\pi}.\label{eq:20}
\end{eqnarray}
\section{Numerical Results and Discussion}
With $d_1=\frac{4}{5}g_A$ fixed from Ref.\cite{Li:2025ddx}, the masses of the decuplet pentaquarks are governed by two LECs, $m_0$ and $f_1$, both of which are determined by fitting to the available experimental data. We take $P_{\psi}^{N}(4312)^{+}$, $P_{\psi s}^{\Lambda}(4338)^{0}$, and $P_{\psi s}^{\Lambda}(4459)^{0}$ all with spin-parity $J^{P}=\frac{1}{2}^{-}$, and use two pairs, $\left(P_{\psi}^{N}(4312)^{+}, P_{\psi s}^{\Lambda}(4338)^{0}\right)$ and $\left(P_{\psi}^{N}(4312)^{+}, P_{\psi s}^{\Lambda}(4459)^{0}\right)$, as inputs.

In our previous work, we took $P_\psi^N(4312)$ and $P_{\psi s}^\Lambda(4338)$ as members of the pentaquark octet as inputs. In the present work, we treat them as members of the pentaquark decuplet as inputs, take $m_{P_{\psi}^{\Delta}}=4.312\ \text{GeV}$ and $m_{P_{\psi {s}}^{\Sigma}}=4.338\ \text{GeV}$, and obtain:

\begin{eqnarray}
	m_0 - 0.039 f_1 - 0.112 &= 4.312, \label{eq:21}
\end{eqnarray}
\begin{eqnarray}
	m_0 - 0.338 f_1 - 0.175 &= 4.338. \label{eq:22}
\end{eqnarray}
Solving this system, we get

\begin{eqnarray}
	m_0 &= 4.413\ \text{GeV}, \label{eq:23}
\end{eqnarray}
\begin{eqnarray}
	f_1 &= -0.3.\label{eq:24}
\end{eqnarray}
Using the obtained values of $m_0$ and $f_1$, we predict the masses of the other two types of hidden-charm pentaquark states $m_{P^{N}_{\psi {ss}}}$ and $m_{P^{\Lambda}_{\psi {sss}}}$:
\begin{eqnarray}
	m_{P^{N}_{\psi {ss}}} &= 4.365\ \text{GeV}, \label{eq:25}\\
	m_{P^{\Lambda}_{\psi {sss}}} &= 4.392\ \text{GeV}.\label{eq:26}
\end{eqnarray}

When we take $P_\psi^N(4312)$ and $P_{\psi s}^\Lambda(4459)$ as inputs with $m_{P_\psi^{\Delta}}=4.312\ \text{GeV}$ and $m_{P_{\psi s}^{\Sigma}}=4.459\ \text{GeV}$, we obtain:
\begin{eqnarray}
	 m_0 - 0.039 f_1 - 0.112   &= 4.312, \label{eq:27}
\end{eqnarray}
\begin{eqnarray}
	 m_0 - 0.338 f_1 - 0.175  &= 4.459. \label{eq:28}
\end{eqnarray}
Solving this system, we get
\begin{eqnarray}
	m_0 &= 4.397\ \text{GeV}, \label{eq:29}
\end{eqnarray}
\begin{eqnarray}	
	f_1 &= -0.7. \label{eq:30}
\end{eqnarray}
Using the obtained values of $m_0$ and $f_1$, we predict the masses of the other two types of hidden-charm pentaquark states $m_{P^{N}_{\psi {ss}}}$ and $m_{P^{\Lambda}_{\psi {sss}}}$:
\begin{eqnarray}
	m_{P^{N}_{\psi {ss}}} &= 4.607\ \text{GeV}, \label{eq:31}\\
	m_{P^{\Lambda}_{\psi {sss}}} &= 4.755\ \text{GeV}.\label{eq:32}
\end{eqnarray}
Interestingly, in both sets of results, the mass splittings between adjacent strangeness layers are nearly constant. This equal-spacing pattern is not accidental; it has a firm theoretical basis in the GMO mass relation, which we now discuss.

The GMO mass relation is one of the earliest and most successful predictions of the SU(3) flavor symmetry. In 1961, Gell-Mann proposed the " eightfold way" framework \cite{Gell-Mann:1962yej}, organizing the observed baryons and mesons into SU(3) irreducible representations. Independently, Okubo derived a general mass-splitting formula for states within a given SU(3) irreducible representation \cite{Okubo:1961jc}. The formula, valid to leading order in the SU(3)-breaking perturbation, reads:
\begin{equation}
	M = a + bS + c\left[I(I+1) - \frac{1}{4}S^2\right],
\end{equation}
where $S$ is the strangeness, $I$ is the isospin, and $a, b, c$ are constants that depend on the representation but not on the individual quantum numbers of the members.

For the baryon octet ($N, \Lambda, \Sigma, \Xi$), substituting the quantum numbers into the GMO formula yields the well-known octet relation \cite{Gell-Mann:1962yej}\cite{Okubo:1961jc}:
\begin{equation}
	\frac{3}{4}M_\Lambda + \frac{1}{4}M_\Sigma = \frac{1}{2}(M_N + M_\Xi),
\end{equation}
which is satisfied by the experimental masses to within about 2\% accuracy.

For the baryon decuplet ($\Delta, \Sigma^*, \Xi^*, \Omega$), the situation is particularly simple. The decuplet members have $I=3/2, 1, 1/2, 0$ and $S=0, -1, -2, -3$, respectively. Substituting these quantum numbers into the GMO formula, the $I(I+1)-S^2/4$ term happens to be a constant across the decuplet, leaving the mass as a linear function of $S$ alone. This leads directly to the equal-spacing rule \cite{Gell-Mann:1962yej}:
\begin{equation}
	M_{\Sigma^*}-M_\Delta \approx M_{\Xi^*}-M_{\Sigma^*} \approx M_\Omega-M_{\Xi^*},
\end{equation}
which was famously confirmed by the discovery of the $\Omega^-$ hyperon in 1964.

In our NLO chiral perturbation calculation, we can examine whether this equal-spacing pattern survives beyond the leading order. The contact terms (proportional to $f_1$) indeed preserve the equal-spacing rule exactly: the three adjacent mass splittings all carry the same coefficient $0.299\,f_1$, as their flavor structure is dictated by the leading-order SU(3)-breaking pattern. The one-loop corrections (proportional to $d_1^2$), however, introduce a small deviation. The three adjacent loop contributions are $0.0593\,d_1^2$, $0.0601\,d_1^2$, and $0.0609\,d_1^2$, respectively, corresponding to a relative deviation of about 2.6\%. This small deviation arises because the Goldstone boson masses ($M_\pi, M_K, M_\eta$) themselves do not satisfy the GMO relation exactly. Nevertheless, the overall equal-spacing pattern remains intact at the NLO, confirming that the GMO relation is robust under chiral loop corrections in the hidden-charm pentaquark system.

Since the constituent quark model predicts a strange--light quark mass difference of approximately 150 MeV, we favor the second set of results as our main predictions. In our results, the mass increment per additional strange quark is about 0.147--0.149 GeV, in good agreement with this expectation.

\section{Summary and Outlook}

We have calculated the NLO chiral mass corrections for the hidden-charm decuplet pentaquarks. The mass spectrum of the full $J^P=\frac{1}{2}^-$ decuplet has been predicted using the experimentally observed states as inputs.

The GMO equal-spacing relation holds at leading order of SU(3) flavor symmetry breaking. Our results show that it remains valid at the next-to-leading order. The tree-level terms satisfy this relation by construction, while the one-loop corrections shift the adjacent mass splittings only slightly. This confirms that the chiral loop effects are modest in the hidden-charm pentaquark system and do not disrupt the leading-order mass pattern.

Several open questions remain. First, the present calculation is restricted to the $J^P=\frac{1}{2}^-$ decuplet, while the $J^P=\frac{3}{2}^-$ decuplet has not been investigated yet. It would be interesting to extend the formalism to the spin-3/2 case and systematically compare the mass splitting patterns between different spin-parity states. Second, the calculation has been carried out only for the hidden-charm sector; whether the same equal-spacing pattern persists in the hidden-bottom decuplet remains an open question. Finally, we note that the experimental identification of whether the observed states belong to the octet or the decuplet is still not definitive, and future experimental and theoretical efforts are needed to clarify the flavor representation of the newly observed hidden-charm pentaquarks.

\end{document}